\documentclass[fleqn,10pt]{wlpeerj} 

\usepackage[hyphens]{url}
\usepackage[multiple]{footmisc} 
\usepackage{pgfplots}
\usepackage{wrapfig} 
\usepackage[font=small,justification=justified,textfont=it]{caption} 

\title{Vocal Interactivity in Crowds, Flocks and Swarms: Implications for Voice User Interfaces}

\author{Roger K. Moore}
\affil{Speech \& Hearing Research Group, Computer Science, University of Sheffield, UK}

\begin{abstract}
Recent years have seen an explosion in the availability of Voice User Interfaces.  However, user surveys suggest that there are issues with respect to usability, and it has been hypothesised that contemporary voice-enabled systems are missing crucial behaviours relating to user engagement and vocal interactivity.  However, it is well established that such \emph{ostensive} behaviours are ubiquitous in the animal kingdom, and that vocalisation provides a means through which interaction may be coordinated and managed between individuals and within groups.  Hence, this paper reports results from a study aimed at identifying generic mechanisms that might underpin coordinated collective vocal behaviour with a particular focus on closed-loop negative-feedback control as a powerful regulatory process.  A computer-based real-time simulation of vocal interactivity is described which has provided a number of insights, including the enumeration of a number of key control variables that may be worthy of further investigation.
\end{abstract}

\begin{document}

\flushbottom
\maketitle
\thispagestyle{empty}

\section*{Introduction} \label{sec:INTRO}

\subsection*{Background}

Recent years have seen an explosion in the availability of `voice user interfaces' (VUIs), initially stimulated by the 2011 launch of \emph{Siri} - Apple's smartphone-based voice assistant - followed in 2015 by Amazon's release of the first `smart speaker' - \emph{Alexa}.  Since then, such smartphone and smart speaker based voice assistants have become almost ubiquitous.  For example, \emph{Siri} has had over 40 million monthly active users in the U.S.\ since July 2017, and smart speaker shipments reached 78 million units worldwide in 2018\footnote{\url{https://medium.com/swlh/the-past-present-and-future-of-speech-recognition-technology-cf13c179aaf}}\footnote{\url{https://www.canalys.com/newsroom/smart-speaker-market-booms-in-2018-driven-by-google-alibaba-and-xiaomi}}.  In the U.K., the number of people who own a smart speaker doubled from one-in-twenty to one-in-ten over a period of just six-months from autumn 2017 to spring 2018\footnote{\url{https://yougov.co.uk/topics/politics/articles-reports/2018/04/19/smart-speaker-ownership-doubles-six-months}}.

However, setting aside the impressive sales figures, a more critical aspect of such voice assistants is the extent to which they are actually used.  For example, a survey conducted in 2015 (i.e.\ prior to the appearance of the first smart speaker) found that only 26\% of the respondents used a voice assistant regularly and the majority of voice assistant users preferred typing to talking \citep{Moore2016a}.  A more recent study by \cite{Kim2019} investigating the usage of voice assistants on both smartphones and smart speakers found that over half of the smart speaker owners used their voice assistant several times a day.  In contrast, only one-third of smartphone owners used their voice assistants on a daily basis, and half hardly used their voice assistants at all.

These studies also reveal that the majority of users employ quite stylised language, e.g.\ using simple voice commands to access music playlists, to perform searches using spoken queries, or to set alerts and reminders.  Such shallow linguistic interaction is somewhat predictable given the nature of the problems users encounter with contemporary voice-enabled devices.  For example, nearly half of the users surveyed reported difficulties with not being understood or simply not being able to do very much.

\subsection*{Key challenges}

The usage statistics reported above confirm that contemporary VUIs are still a long way from being able to provide the ``\emph{conversational interface}'' often promoted in the marketing literature for such systems.  Indeed, the fact that users are effectively resorting to `voice button-pressing' suggests that there is a fundamental difference between the richness of everyday human-human spoken language and the simplicity of the voice-based interaction that takes place between humans and machines.  It has been argued elsewhere that a `mismatch' between interlocutors is not only an important obstacle that needs to be explored in a human-machine context \citep{Moore2015}, but that it may even be an unsurmountable problem \citep{Moore2016c}.  In particular, if spoken language interaction is viewed as being based on the co-evolution of two key traits -- \emph{ostensive-inferential} communication and \emph{recursive mind-reading} \citep{Scott-Phillips2015}, then contemporary voice-based systems are essentially only dealing with one aspect -- inference.  Some of the high-level issues relating to recursive mind-reading have been addressed in \cite{Moore2017b}, but low-level concerns relating to ostensive vocal behaviour remain an open question.  Hence, there seems to be something essential missing from contemporary voice-enabled system in the area of user engagement and interaction -- not just what to say, but when to say it and, in a group context, to whom \citep{Moore2015}.

\subsection*{Potential solutions}

Of course, interactive ostensive behaviours are ubiquitous in the animal kingdom, and vocalisation provides a means through which such activities may be coordinated and managed between individuals and within groups \citep{Moore2016b}.  Vocalisations are often carefully timed in relation to each other (and other events taking place in an environment), and this may take the form of \emph{synchronised} ritualistic behaviour (such as rhythmic chanting, chorusing or singing) or \emph{antisynchronous} turn-taking (which can be seen as a form of dialogue) \citep{Cummins2014,Fusaroli2014,Ravignani2014}.

Of particular interest here are the \emph{mechanisms} that support the emergence of synchronised vocal interactivity in crowds, flocks and swarms, and the implications of those mechanisms for future voice user interfaces.  Hence, this paper presents results from an investigation into such mechanisms using a real-time simulation (i.e.\ a computational model) of interactive vocal dynamics and synchrony.

\section*{Collective Behaviour}

The coordinated behaviour of large numbers of independent living organisms has been the subject of scientific enquiry for many years.  For example, studies have been conducted into the flocking of birds \citep{Reynolds1987a}, the synchronised flashing of fireflies \citep{Ermentrout1991}, the dynamics of human crowd movement \citep{Still2000}, waves of coordinated clapping by audiences \citep{Neda2000}, and spatial sorting in shoals of fish \citep{Couzin2002}.  Of particular interest are the transitions from one type of collective behaviour to another, especially in the context of attraction and repulsion between individuals \citep{Katz2011}, and predator-prey interactions \citep{Handegard2012}.  Much of the research has involved computational simulation (perhaps the most famous being `\emph{Boids}'\footnote{\url{https://www.red3d.com/cwr/boids/}}), as well as physical implementations in the field of swarm robotics, e.g.\ \cite{Bo2005}.

One important aspect of synchronous behaviours is that they involve parallel coupled simultaneous action, as opposed to sequential action-reaction \citep{Cummins2011}.  Such collective behaviours can thus be viewed as rhythmic entrainment, and thereby constitute a form of accommodation between individuals in a population \citep{DeLooze2014}.  It has also been posited that such behaviours underpin the links between different modalities, such as between vocalisation and physical movement \citep{Cummins2009a}.

\subsection*{Vocal Synchrony}

Whilst there have been a number of studies of vocal synchrony in animals, e.g.\ male zebra finches \citep{Benichov2016} and monkeys \citep{Takahashi2013}, what is important here is the synergy with human vocal behaviours.  Much of this work has involved `joint speech', i.e.\ people speaking in unison \citep{Cummins2014}, and a more sophisticated view of `turn-taking' in human dialogue \citep{Heldner2010}.  Of particular relevance is evidence that verbal synchrony in large groups of people produces affiliation \citep{VonZimmermann2016}, and that some conversational partners tend to converge their vocal behaviours \citep{Edlund2009}, while others do not \citep{Assaneo2019}.

\subsection*{Mechanisms}

With regard to the mechanisms underpinning coordinated collective behaviour, by far the most popular approach is based on \emph{coupled oscillators}  \citep{Kuramoto1975,Strogatz1993,Strogatz2012}, particularly through `pulsatile coupling' \citep{Mirollo1990}.  Not only has this been a very productive modelling paradigm with real-world implications (such as the simulation of clustered synchrony in electricity distribution networks \citep{Pecora2014}), but new results are continuing to emerge \citep{Matheny2019}.  The coupled-oscillator paradigm is also attractive because of its potential compatibility with known neural mechanisms \citep{Ermentrout1991,Matell2000}.  However, it is only one way of formulating a complex non-linear attractor space, and it overlooks a number of potentially important conditioning variables -- e.g.\ \emph{energetics} \citep{Moore2012i}.

As a consequence, the work reported here departs from the standard coupled-oscillator approach.  In particular, attention is given to an alternative paradigm for creating a space of behavioural attractors -- `closed-loop negative-feedback control' -- a powerful \emph{regulatory} mechanism with roots in `cybernetics' \citep{Wiener1948} and commonly deployed for stabilising engineering systems \citep{DiStefanoIII1990} as well as providing a powerful \emph{non-behaviourist} paradigm for modelling the behaviour of living systems \citep{Powers1973}.  The main differences between this approach and coupled oscillators is that the convergence criteria can be made more explicit, thereby offering the potential to gain a deeper understanding of the implications of particular parameters/settings on the emergent collective behaviours.  It also offers the advantage that it can, in principle, be generalised to the synchronisation of more complex metrical structures, e.g.\ as discussed by \cite{Fitch2013}.

\section*{Simulation Framework} \label{sec:SIM}

\subsection*{Basic principles}

The basic operation of classic closed-loop negative-feedback control is as follows: (i) a reference signal specifies the \emph{desired} consequences of a system's actions, (ii) the \emph{actual} consequences are sensed/interpreted by the system and compared with the reference target, (iii) the resulting \emph{error} generates a control signal that drives the system in a direction such that the error is minimised.  The process continues around the loop causing the system to not only converge to the desired behaviour but, more significantly, to maintain that behaviour in the face of arbitrary disturbances \emph{without} having to sense such disturbances directly.

The tracking behaviour of such a negative-feedback control system is a function of the `loop gain' of the controller.  If the loop gain is too low, then stabilisation may take a long time -- an `overdamped' system.  On the other hand, if the loop gain is too high, then the system may overshoot and even oscillate -- an `underdamped' system.  The point here is that the loop gain effectively corresponds to the degree of \emph{effort} (energy) applied to a regulatory task, i.e.\ from a psychological standpoint, it is analogous to \emph{motivation}.  An agent that `cares' about controlling a particular variable would have a high loop gain, whereas a loop gain of zero implies the agent doesn't care at all (i.e.\ it gives up control).  These are important individual differences that are not explicit in the coupled-oscillator approach.

\subsection*{Implementation}

The simulations described herein have been implemented in Pure Data\footnote{\url{http://puredata.info/}} -- known as ``\emph{Pd}'' -- an open-source object-oriented dataflow programming language that is designed for real-time audio processing \citep{Farnell2008}.  An environment has been constructed in which any number of vocalising (and listening) `agents' may be connected to each other in arbitrary network topologies.  Each agent comprises two feedback-control loops: one to regulate the interaction with other agents and another to regulate the agent's own behaviour.  The first of these control loops aims to maintain synchrony between an agent's own vocalisations and those from agents that it can `hear' (i.e.\ those to which it is connected).  The second control loop attempts to maintain the agent's own preferred vocal rhythm.  This arrangement means that each agent has two control parameters that influence the priority given to `self' versus `other'.  

In addition, each agent has settings for the amplitude and duration of its vocalisations, their phase relation with the rhythmic `beat', and the agent's preferred rhythm.  In principle, these parameters could also be the subject of optimisation using feedback-control, but this was not implemented in the experiments reported here.  

The sound output from each agent was produced using real-time synthesis of human, bird or insect vocalisations (as selected by the user).  Other vocal characteristics for each agent (e.g.\ pitch frequency) were initialised randomly in order to provide a moderate level of `individuality'.  

\begin{figure}[h!]
  \centering
  \includegraphics[width=0.9\linewidth]{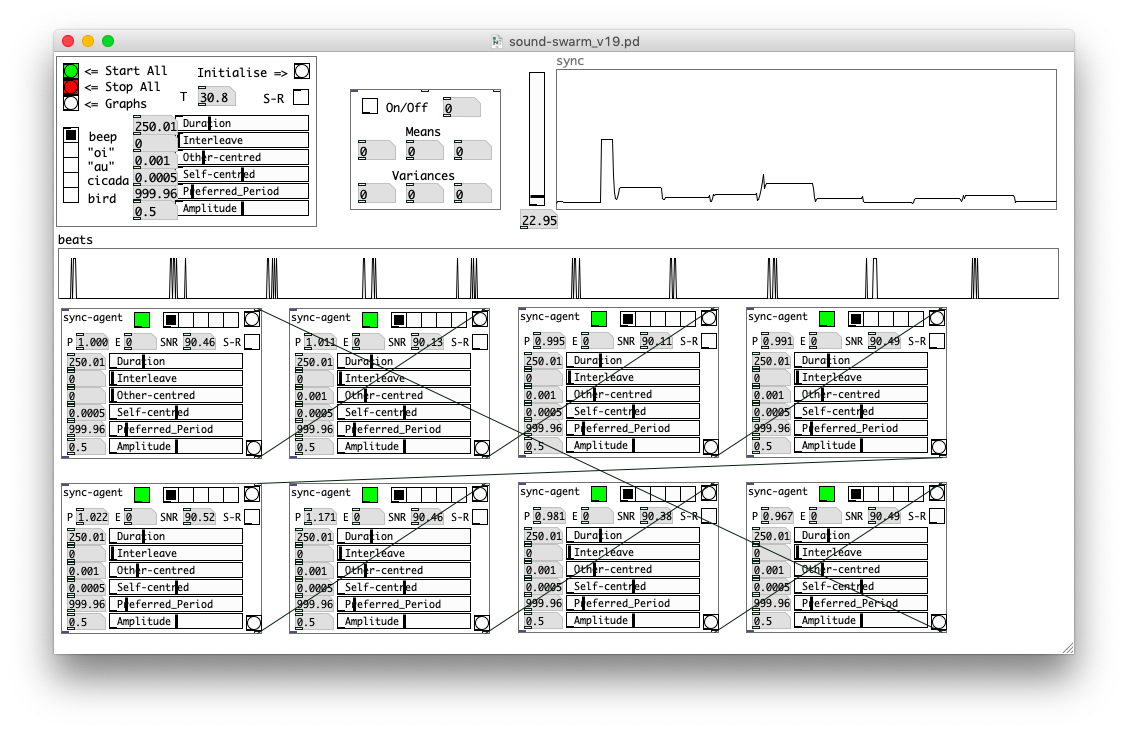}
  \caption{Screenshot of the user interface for the \emph{Pd}-based vocal simulation environment.  The main control panel is shown in the top-left corner; buttons and sliders allow a user to specify global parameters for the population of agents, such as the type of vocalisation (human, insect, bird), duration, loudness etc.  The lower half of the user interface facilitates the creation of an arbitrary number of agents, and the specification of which agent is listening to which other agent(s).  In the example shown, eight agents have been configured in a loop topology (agent-2 is listening to agent-1, agent-3 to agent-2, agent-4 to agent-3, \ldots, agent-1 to agent-8).  As can be seen, sliders on each agent allow a user to set parameters individually if required.  The graph shown at the top-right of the interface displays a timeline of the overall vocal synchrony between the agents, and the graph shown across the middle displays the individual rhythmic `beats' from each agent (bunching indicates a degree of synchrony).}
  \label{fig:SS19-8}
\end{figure}

Figure~\ref{fig:SS19-8} illustrates a particular configuration of the simulation environment.

\subsection*{Experiments}

A range of experiments has been conducted based on varying numbers of interacting agents, different interconnection topologies, and alternative parameter settings.  There is not space here to report all the findings.  So what follows is a selected highlight.

\begin{wrapfigure}{r}{0.5\linewidth}
	\centering
\scalebox{0.7}{
	\begin{tikzpicture}
		\begin{axis}[xlabel=Agent's Position in Chain,ylabel=Synchronisation Error (msecs),legend style={at={(0.05,0.85)},anchor=west}]
			\addplot+[black,mark options={fill=black},error bars/.cd,
			y dir=both,y explicit]
			coordinates {
			(2,-0.58) +- (3.007,3.007)
			(3,-2.05) +- (7.874,7.874)
			(4,-4.72) +- (20.43,20.43)
			(5,-4.33) +- (37.93,37.93)
			(6,-8.1) +- (84.51,84.51)
			(7,-12.6) +- (132.6,132.6)
			(8,-13.7) +- (134.5,134.5)
			};
			\addplot+[black,mark options={fill=black},error bars/.cd,
			y dir=both,y explicit]
			coordinates {
			(2,-23.8) +- (0.185,0.185)
			(3,-47.6) +- (0.371,0.371)
			(4,-71.4) +- (0.418,0.418)
			(5,-95.3) +- (0.221,0.221)
			(6,-119) +- (0.495,0.495)
			(7,-142) +- (0.344,0.344)
			(8,-166) +- (0.959,0.959)
			};
			\legend{feedback-control, action-reaction}
		\end{axis}	
	\end{tikzpicture}
	}
	\caption{Relationship between a `slave' agent's position in a chain and its vocal synchronisation error with respect to the `master' agent at the head of the chain.}
	\label{fig:EXP}
\end{wrapfigure}
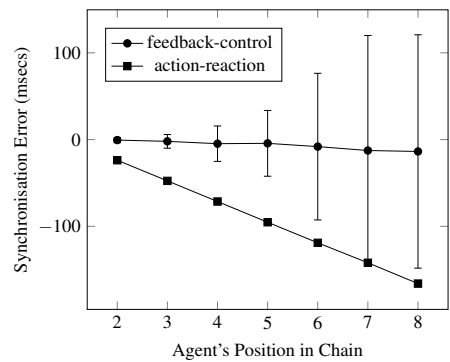

One of the overarching research questions is concerned with the relationship between the topological connections between agents (i.e.\ the `ostensive' relationships) and the emergent collective behaviour of the agents.  In this context, one particular configuration is a \emph{chain} with a `master' (pacemaker) agent and a sequence of `slave' agents.  Figure~\ref{fig:EXP} illustrates the outcome of simulating such a configuration with a chain of eight agents.  As can be seen, on average, all of the agents in the feedback-control configuration maintained synchrony, but the agents further down the chain exhibited less stable rhythms.  In contrast, agents in an action-reaction configuration maintained the rhythm, but the agents further down the chain were increasingly out of sync.

\section*{Discussion \& Conclusion}

As a result of this research, it is possible to draw some conclusions about the control variables that are worthy of investigation with respect to vocal interactivity.  These are summarised in Table~\ref{tab:DIM}.

\begin{table}[h!]
	\caption{Dimensions of vocal interactivity identified in this study.  The left-hand column specifies the relevant control variables, and the right-hand column gives an indication of the expected range of values.}
	\label{tab:DIM}
	\centering
	\footnotesize
	\begin{tabular}{| l | c |}
		\hline
		\multicolumn{2}{| c |}{\textbf{VOCAL AGENTS}} \\ \hline
		Individuality (e.g.\ style of vocalisation) & average $\Leftrightarrow$ extreme \\ \hline
		Ostention (i.e.\ stance towards other agents) & connected $\Leftrightarrow$ disconnected \\ \hline
		Intentionality (i.e.\ goals wrt other agents) & same $\Leftrightarrow$ different \\ \hline
		Motivation/effort expended on pursuing others' goals & high $\Leftrightarrow$ low \\ \hline
		Motivation/effort expended on pursuing own goals & high $\Leftrightarrow$ low \\ \hline\hline
		\multicolumn{2}{| c |}{\textbf{VOCALISATIONS}} \\ \hline
		Intensity (e.g.\ volume) & low $\Leftrightarrow$ high \\ \hline
		Clarity (e.g.\ intelligibility/SNR) & low $\Leftrightarrow$ high \\ \hline
		Period (i.e.\ timing) & short $\Leftrightarrow$ long \\ \hline
		Mark-to-space ratio (i.e.\ duration) & 0\% $\Leftrightarrow$ 100\% \\ \hline
		Sentiment (i.e.\ valence) & -ve $\Leftrightarrow$ +ve \\ \hline
		Meaning (e.g.\ category) & named-entity-1 $\Leftrightarrow$ named-entity-2 \\ \hline\hline
		\multicolumn{2}{| c |}{\textbf{VOCAL INTERACTVITY}} \\ \hline
		Synchrony (i.e.\ engagement) & low $\Leftrightarrow$ high \\ \hline
		Simultaneity (i.e.\ overlap/interleaving) & 0\% $\Leftrightarrow$ 100\% \\ \hline
		Dependency (i.e.\ between vocalisations) & dependent $\Leftrightarrow$ independent \\ \hline
	\end{tabular}
  \end{table}

In conclusion, this paper has outlined some of the key issues facing contemporary voice-user interfaces, with a special focus on emergent properties of collective vocal behaviour, especially ostensive interaction and timing.  The focus has been on closed-loop negative-feedback control as a regulatory mechanism, which implements a coincidence detection scheme that is compatible with known neural mechanisms \citep{Matell2000}.  The simulation of real-time interacting vocal agents has already provided a number of insights into such behaviour, and more are expected as the full parameter space is investigated.  In particular, it should be possible to show (i) how dialogue emerges as a compensatory response to the automatic regulation of intelligibility, not as a trivial action-reaction behaviour \citep{Benichov2016}, (ii) how cooperative \emph{vs.} competitive interaction conditions vocalisations, and (iii) how communicative behaviour emerges from vocal interaction \citep{Rosenthal2015}.

\begin{small}
\bibliography{mybib}
\end{small}

\end{document}